\documentclass[aps,prl,reprint,showpacs,floatfix,shortbibliography,raggedbottom]{revtex4-2}

\draft 
\usepackage[utf8]{inputenc}
\usepackage{amsmath,amsthm,amssymb}
\usepackage{float}
\usepackage{graphicx}
\usepackage{dcolumn}
\usepackage{bm}
\usepackage{color}
\usepackage{soul}
\usepackage{cancel}

\usepackage{hyperref}   
\usepackage[all]{hypcap}
\pdfoutput=1
\DeclareUnicodeCharacter{2212}{-}

\begin{document}
\preprint{APS/123-QED}

\title{Continuum damping of topologically-protected edge modes at the boundary of cold magnetized plasmas}
\author{Roopendra Singh Rajawat}
\email{rupn999@gmail.com}
\affiliation{School of Applied and Engineering Physics, Cornell University, Ithaca, NY 14850}
\author{V. Khudik}
\affiliation{Department of Physics and Institute for Fusion studies, The University of Texas at Austin, TX 78712}
\author{G. Shvets}
\affiliation{School of Applied and Engineering Physics, Cornell University, Ithaca, NY 14850}
\date{\today}

\begin{abstract}
Recent extension of the topological ideas to continuous systems with broken time-reversal symmetry, such as magnetized plasmas, provides new insights into the nature of scattering-free topologically-protected surface plasma waves (TSPWs). We demonstrate a unique characteristic of TSPWs propagating above the electron cyclotron frequency: their collisionless damping via coupling to the continuum of resonant modes localized inside a smooth plasma-vacuum interface. Damped TSPWs retain their unidirectional nature and robustness against backscattering. When sheared magnetic field creates a boundary between damped and undamped TSPWs, the two refract into each other without reflections.
\end{abstract}
\maketitle

{\it Introduction.-}
Independently of their state of matter -- solid~\cite{Hasan_rmp_2010,Haldane_prl_2008,soljacic_prl_2008}, gaseous and fluid~\cite{delplace_science_2017,Yang_prl_2015}, or plasma~\cite{Parker_prl_2020b,fu_nc_2021} -- a wide range of materials is found to exhibit non-trivial topological properties that fundamentally impact wave propagation at domain walls between topologically-distinct bulk materials. Specifically, bulk materials possessing propagation bandgaps (i.e., acting as ``insulators" for wave-like polaritonic perturbations of certain energies) and lacking the time-reversal (TR) symmetry can be characterized by an integer invariant, known as the Chern number~\cite{Ozawa_rmp_2019}, assigned to every bandgap. The bulk-edge correspondence (BEC) principle  -- originally established in condensed matter physics~\cite{Klitzing_prl_1980,Hatsugai_prl93,Hasan_rmp_2010} and later extended to photonics and metamaterials, cold atomic gases, and classical fluids \cite{Haldane_prl_2008,soljacic_prl_2008,Silveirinha_prb_2015,gangaraj_prl20,Goldman_np_2016,tauber_jfm_2019}, -- predicts the existence and number of gapless unidirectional edge states that are spectrally-located within a common bandgap of the two bulk materials sharing a domain wall.

\begin{figure}[h!]
    \centering
 \includegraphics[width=1.\linewidth]{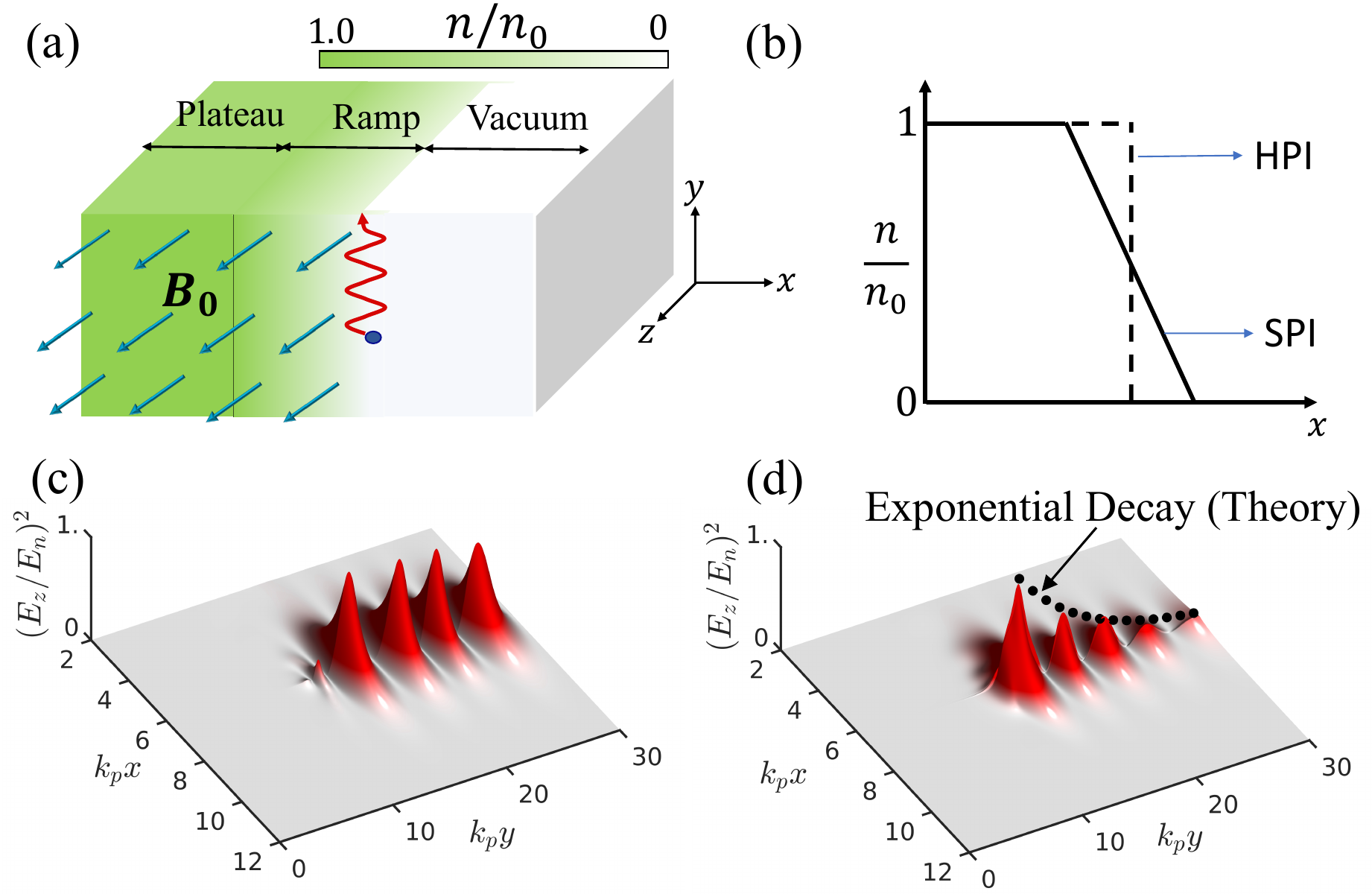}
        \caption{\label{fig:Fig1} (a) Surface wave excitation at the interface between a magnetized plasma slab (green) and vacuum by a dipole source (blue disk) oscillating with $\omega_{\rm dr} = 0.9 \omega_{p0}$. Generated TSPW (red wavy arrow) propagates unidirectionally along the interface.  The source is localized at $y = 12.5 k_p$ (b) Plasma density profiles for hard (dashed line) and smooth (solid line) plasma interfaces. (c,d) Snapshots of time-averaged $E_z^2$ in a $z={\rm const}$-plane for (c) HPI, and (d) SPI. Parameters: $\omega_c/\omega_{p0} = 0.75$, $k_z/k_p = 1.0$, and $k_p \delta l = 0.2$ for SPI.}
\end{figure}

There are several remarkable properties of the topological edge states stemming from BEC. (i) In the real space, they are immune to backscattering with respect to various domain wall perturbations, e.g., changes along (e.g., sharp turns) and across (e.g., interface smoothness) its perimeter. (ii) In the reciprocal space, their dispersion curves connect the two continua of the bulk modes (CBM) by spanning the corresponding topological bandgap. Property (i) is essential for a variety of practical applications, such as plasma-based circulators and isolators~\cite{sievenpiper_prapp_2022}. While both properties are realized in periodic systems, where the band structure is determined by its lattice, the situation can be quite different for continuous systems, such as moving/rotating fluids or magnetized plasmas \cite{Parker_prl_2020b,case_PoF60,briggs_PoF70}, where intrinsic material resonances (e.g., electron cyclotron resonance) create multiple continuous propagation bands. Recent work identified several limitations in applying the BEC to cold magnetized plasmas -- the subject of this Letter -- that prevent the edge states from crossing the entire bulk bandgap\cite{Gangaraj_prl_2020, Han_prb_2022}. For example, air/plasma domain walls comprising nearly-discontinuous density jumps support spectrally-flat short-wavelength surface-plasmon resonances~\cite{Gangaraj_prl_2020}; those limit the spectral range of the topological edge states in violation of the conventional BEC. While a smooth plasma density transition inside the domain wall removes the surface-plasmon resonances ~\cite{Parker_prl_2020b,fu_nc_2021}, it brings to the fore a key distinctive characteristic of bounded continuous media, such as non-uniformly moving/rotating fluids or inhomogeneous plasmas: that a new continuous spectrum of modes can arise in addition to the continuum modes of the bulks ~\cite{case_PoF60,briggs_PoF70,sedlacek_1971,Shvets_pop_1999,Parker_prl_2020a}.

In the specific case of cold magnetized plasmas comprising mobile electrons and immobile neutralizing ions -- where the applied magnetic field breaks the TR symmetry \cite{Parker_prl_2020a} -- the new continuum of modes corresponds to local upper hybrid resonances: $\omega_{\rm UH}^2(\mathbf{x}) = \omega^2_{\rm c} + \omega^2_{\rm p} (\mathbf{x})$, where $\omega_{\rm c} =eB_0/m c$ is the cyclotron frequency of an electron with mass/charge $m/e$ rotating in a uniform magnetic field ${\bf B_0} =  B_0 \hat{z}$, and $\omega_p(\mathbf{x}) = \sqrt{4\pi e^2 n(\mathbf{x})/m}$ is the local plasma frequency inside the domain wall determined by the local unperturbed plasma density $n(\mathbf{x})$. In the simplest case of a semi-infinite volume of cold magnetized plasma with uniform density $n_0$ interfaced with a vacuum region through a smooth domain wall, the continuum of localized modes (CLM) occupies the spectral region $\omega_{\rm c} < \omega < \Omega_{\rm UH}$, where $\Omega_{\rm UH}$ is the upper-hybrid frequency of the bulk.

The presence of the CLM inside the topological bandgap raises a number of fundamental questions that do not arise in the case of a periodic topological medium. For example, it is unclear if topologically robust bandgap-spanning modes -- possibly damped -- can exist inside the CLM spectral band~\cite{Parker_prl_2020b,fu_nc_2021,*Qin_sa_2023}, as one would expect from the BEC. Another open question is how property (i) is affected, i.e., whether the presence of the CLM band affects the robustness of edge state to backscattering when a domain wall makes a sharp turn. It also remains to be understood how the profile of the domain wall affects the damping of the edge states when their frequency overlaps with the continuum of local modes. In this Letter, we use particle-in-cell (PIC) simulations and semi-analytic modeling to demonstrate the existence of a new class of backscattering-immune unidirectional collisionlessly damped modes spanning the entire CLM band, thereby resolving the apparent anomaly of the bulk-edge correspondence. From the standpoint of potential applications, continuum damping can be utilized for depositing the finite energy, momentum, and angular momentum of the TPSWs~\cite{Fu_jpp_2022} into the plasma in a robust way that is not sensitive to the exact shape of the plasma/vacuum interface. Numerous applications of rotating plasmas ranging from isotope separation to improved plasma confinement have been suggested~\cite{rax_pop_2017}.

\begin{figure}[t]
    \centering
    \includegraphics[width=1\linewidth]{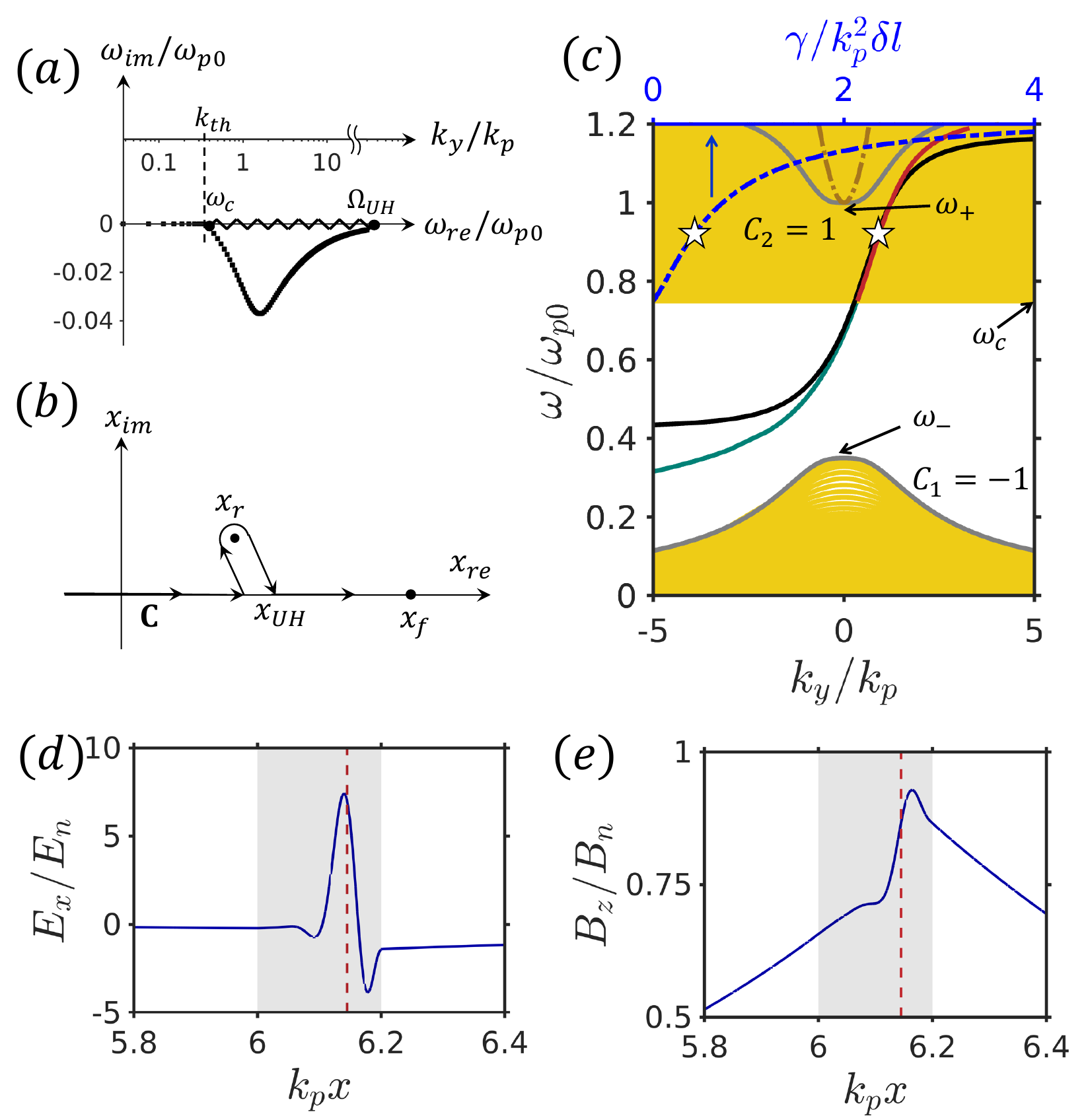}
    \caption{(a,b) Integration contours in the complex (b) $\omega$- and  (c) $x$- planes used for calculating the dispersion relation.  The $k_{th}$ represents threshold value of $k_y$  after that integration contour moves into complex $\omega-$ plane.(c) Propagation band structure for a semi-infinite magnetized plasma slab with a smooth plasma/vacuum interface. Yellow curves: propagating bulk and ramp-localized continuum modes. Undamped surface waves: HPI-TSPW (black solid) and SPI-TSPW (green solid). Damped SPI-TSPWs: spatial propagation ($k_y$: red solid line) and decay ($\gamma$: blue dashed line) constants.  Chern numbers are indicated for the bulk bands. Stars at $\omega = \omega_{\rm dr}$: driven modes from Fig.~\ref{fig:Fig1}(c,d). (d,e) Spatial profiles of (d) $E_{\rm x}$, and (e) $B_{\rm z}$ oscillation amplitudes extracted from the PIC simulation. Shaded region: linear density ramp of the SPI. Red dashed line: upper-hybrid resonance at $x_{\rm UH}$. Plasma parameters: same as in Fig.~\ref{fig:Fig1}.}
    \label{fig:Fig2}
\end{figure}

As a model for investigating topological surface plasma waves (TSPWs), we use a physical configuration shown in Fig.~\ref{fig:Fig1}(a): a planar slab of magnetized plasma with density $n(\mathbf{x}) \equiv n(x)$ (where $n(x) \rightarrow n_0$ for $x \ll L_{\rm int}$) interfacing with a vacuum region at $x \approx L_{\rm int}$. Two types of interfaces shown in Fig.~\ref{fig:Fig1}(a) are utilized: a hard plasma interface (HPI: solid line) and a smooth plasma interface (SPI: dashed line). To excite TSPWs with a fixed axial wavenumber $k_z$, we use a periodic chain of oscillating point sources with dipole moments $\mathbf{p} \equiv \mathbf{e}_z p_z$ (blue dot in Fig. \ref{fig:Fig1}(a)) oscillating as $p_z \propto \cos \left( k_z \cdot z - \omega t \right)$ spaced by $P_z \ll 2\pi/k_z$ in the $z-$direction placed near the plasma-vacuum interface.

The excitation of TSPW's with $k_z = \omega_{p0}/c$ propagating along with the interface ($\pm yz-$plane) was modeled by a first-principles 3D-PIC simulation code VLPL\cite{pukhov_1999} for cold collisionless plasmas with magnetic field corresponding to $\omega_c = 0.75 \omega_{p0}$: see~\cite{supplemental} for the details of the PIC simulation setup. For these parameters, the upper (lower) edge of the first (second) bulk propagation band with the Chern number $C_1=-1$ ($C_2=1$) is at $\omega_{-} \approx 0.4 \omega_{p0}$ ($\omega_{+} \approx \omega_{p0}$), resulting in a complete bandgap for $\omega_{-} < \omega < \omega_{+}$. For the smooth plasma interface, our driven PIC simulations reveal the existence of undamped TSPWs in the lower portion of the bandgap: $\omega_{-} < \omega < \omega_{\rm c}$, as was recently demonstrated using eigenvalue simulations~\cite{Parker_prl_2020b,fu_nc_2021}. The dispersion relation $\omega(k_y)$ for these modes is plotted as a green line in Fig. \ref{fig:Fig2}(c) for an SPI represented by a linear density ramp with the length $\delta l = 0.2 k_p^{-1}$ (where $k_p^{-1} \equiv c/\omega_{p0}$ is the collisionless skin depth of the bulk plasma). Eigenvalue simulations do not reveal any edge states -- damped or undamped -- in the rest of the bulk bandgap overlapping with the CLM band extending from $\omega_{\rm c}$ to $\Omega_{\rm UH} > \omega_{+}$.

However, driven PIC simulations at $\omega_{\rm dr} = 0.9\omega_{p0}$ for the same SPI (Fig. \ref{fig:Fig1}(d)) and for the HPI (Fig. \ref{fig:Fig1}(c): $\delta l = 0$) present a different picture. The red-colored images of the time-averaged ${E}_z^2$ reveal that the edge states are indeed launched inside the CLM band for both interface types, and illustrate several of their properties. First, SPI and HPI support unidirectional edge states propagating in the positive $y-$direction. Second, while the HPI-TSPW mode propagates undamped with $k_y \approx 0.8k_p$, the SPI-TSPW appears to be exponentially decaying, with the corresponding complex-valued wavenumber $\tilde{k}_y \approx \left( 0.8 - 0.075i \right) k_p$. The lack of damping in the HPI case is not surprising because the CLM band collapses into a single point at $\omega = \Omega_{\rm UH}$ outside of the bulk bandgap. However, the presence of weakly-damped modes in the SPI case is puzzling because no such modes are revealed by eigenvalue simulations. The collisionless nature of our PIC simulations raises the question of how the energy of TSPWs is dissipated in the absence of plasma viscosity.

Nevertheless, several conjectures -- validated below -- can be made and utilized for interpreting the physics of the collisionlessly-damped unidirectional TSPWs. First, apart from finite damping, TSPW propagation along relatively sharp ($\delta l \ll k_p^-1$) interfaces appears similar to their propagation along the HPI: $\tilde{k}_y \approx k_y$. Therefore, a perturbed solution for SPI-TSPWs can be obtained using the field profile of HPI-TSPWs as the unperturbed solution~\cite{Pakniyat_ieee_2020}, and $k_p \delta l$ as a small parameter. Second, the unidirectional nature of the SPI-TSPWs indicates that they are robust to backscattering by abrupt (i.e., on a scale shorter than $\Delta y \sim k_y^{-1}$) perturbations of the domain wall. Third, collisionless damping of SPI-TSPWs appears to be a kinetic effect that results from strong plasma heating occurring at the specific location $x=x_r$ such that the unperturbed HPI-TSPW frequency $\omega$ resonates with highly-localized upper-hybrid resonances: $\omega_{\rm UH}(x_r) = \omega$. Resonant excitation of localized upper-hybrid waves can produce energetic electrons \cite{Lin_pof_1982}.

Finally, the decaying mode SPI-TSPW is highly reminiscent of weakly-damped surface waves in un-magnetized plasmas that were interpreted as quasi-modes generated by phase mixing of multiple local resonances~\cite{Shvets_pop_1999}. The key distinction between the true exponentially-decaying eigenmodes (e.g., due to finite fluid viscosity) and the quasi-modes is that the latter exhibit short-term exponential and long-term power-law decay that can be viewed as a form of Landau damping~\cite{landau_1946} in physical space~\cite{briggs_PoF70,strogatz_prl92}. While continuum damping of electromagnetic waves is not unusual in plasma physics~\cite{sedlacek_1971,MHD_Book}, to our knowledge, it has never been studied in the context of topologically-robust surface waves in magnetized plasmas.

To construct a cold fluid description of the edge states, we assume that their electric ($\bf E$) and magnetic ($\bf B$) fields are harmonic, i.e. proportional to $e^{i \mathbf{k} \cdot \mathbf{x} - i \omega t}$, where $\mathbf{k} = \mathbf{e}_y  k_y + \mathbf{e}_z  k_z$ is the in-plane propagation wavenumber.  Temporal or spatial damping rates of TSPWs are included in their complex-valued frequency ($\omega$) or wavenumber ($k_y$), respectively~\cite{Archambault_prb_2009}. After space-time Fourier transformation of Maxwell equations \cite{Jackson}, we obtain:
\begin{equation}\label{eq:shooting_method}
  \frac{\partial \mathbf{\psi}}{\partial x}   = -\frac{i}{\epsilon_t(x)} M \mathbf{\psi}, \ \ \ \mathbf{\psi}  = \left(B_x,B_y,B_z,E_z\right)^T
\end{equation}
where $\mathbf{\psi}$ is a four-component vector comprising the field components remaining continuous across the discontinuous HPI. The $4 \times 4$ matrix $M\left(x,\mathbf{k},\omega \right)$ is given by
\begin{equation} \label{eq1}
M = \begin{pmatrix}
0 & k_y \epsilon_t & k_z \epsilon_t & 0 \\
-k_y \epsilon_t & 0 & 0 & k_0 \epsilon_a \epsilon_t \\
\frac{k_0^2 \epsilon_{tg}^2 - \epsilon_t k_z^2}{k_z} & -ik_z \epsilon_g & ik_y \epsilon_g & - \frac{k_0 k_y}{k_z}  \epsilon^2_{tg} \\
ik_0 \epsilon_g & \frac{\epsilon_t^2 k_0^2 - k_z^2}{k_0} &  \frac{k_0 k_y}{k_z}  & -ik_y \epsilon_g
\end{pmatrix},
\end{equation}
where $k_0 = \omega/c$, $ \epsilon_{t}  =  \frac{\omega^2 - \omega^{2}_{\rm UH}(x)}{ \omega^{2} - \omega^{2}_{c}}$, $\epsilon_{g}  =  \frac{\omega_{c}}{\omega}  \frac{\omega^{2}_{p}(x)}{\omega^{2}_{c} -\omega^{2}}$, and $\epsilon_{a}  = \frac{\omega^2 - \omega^{2}_{p}(x)}{\omega^2}$ are the $x$-dependent components of the cold plasma dielectric tensor \cite{Stix}, and $\epsilon^2_{tg} \equiv \epsilon^2_t - \epsilon_g^2$. The remaining two components of the electric field are expressed in terms of $\mathbf{\psi}$ components:
\begin{eqnarray}
  \epsilon_t E_x &=& \frac{i k_y k_0 \epsilon_g E_z - k_y k_z B_z}{k_0 k_z} + \frac{k_z}{k_0} B_y - i \frac{k_0}{k_z} \epsilon_g B_x,  \label{eq:Ex} \\
  E_y &=& \left( \frac{k_y}{k_z} \right) E_z - \left( \frac{k_0}{k_z} \right) B_x. \label{eq:ExEy}
\end{eqnarray}

Therefore, by integrating Eq.(\ref{eq:shooting_method}) between $x=x_{\rm in} \rightarrow -\infty$ and $x=x_{\rm out} \rightarrow +\infty$ endpoints (deep inside and outside the plasma) and assuming the vanishing of $\mathbf{\psi} \rightarrow 0$ at the endpoints, a TSPW dispersion relation in the form of $D(\omega,k_y)=0$ can be obtained~\cite{Chu_pop_1994,Shvets_pop_1999} for any SPI.  For the TSPWs in the $\omega_{-} < \omega < \omega_{c}$ frequency range, the $x$-integration of Eq.(\ref{eq:shooting_method}) can be carried out along the real axis because no singularities are encountered for any real-valued $x_{\rm in} < x < x_{\rm out}$ for any plasma density profile $n_0(x)$. The resulting solution of the dispersion relation yields a real-valued (damping-free) dispersion relation $\omega(k_y)$ \citep{fu_nc_2021}  plotted as a green solid line in Fig.~\ref{fig:Fig2}(c).

The situation changes dramatically for the real-valued TSPW frequencies inside the CLM band ($\omega > \omega_c; \, k_y > k_{th}$) because the $1/\epsilon_t(x)$ factor in Eq.(\ref{eq:shooting_method}) acquires a pole singularity at the local upper-hybrid resonance location $x=x_{\rm UH}(\omega)$ defined by $\omega^2 = \omega^2_{c} + \omega^2_p (x_{\rm UH})$. The ambiguity of integrating Eq.(\ref{eq:shooting_method}) across the singularity is resolved by analytically continuing the dispersion function ($D \rightarrow D_{\ast}$) into the lower-half of the complex $\omega$-plane to preserve causality~\cite{briggs_PoF70,strogatz_prl92,Shvets_pop_1999}: see Fig. ~\ref{fig:Fig2}(a). If $D_{\ast}(\omega,k_y)$ possesses a pole $\tilde{\omega}_{\rm QM} \equiv \omega^{\rm re}_{\rm QM} + i \omega^{\rm im}_{\rm QM}$ (where $\omega^{\rm im} < 0$), then its analyticity requires that the integration between $x_{\rm in}$ and $x_{\rm out}$ be carried out along the path $\mathcal{C}$ in the complex-$x$ plane passing below the complex-valued point $x_r$ defined as $\omega_{\rm UH}(x_r) = \tilde{\omega}_{\rm QM}$~\cite{briggs_PoF70,Shvets_pop_1999}: see Fig.~\ref{fig:Fig2}(b).

Iterative application of such integration for different trial frequencies $\tilde{\omega}_{\rm tr}$ yields the quasi-mode dispersion relation $\tilde{\omega}_{\rm QM}(k_y)$ satisfying the condition of $D_{\ast}(\tilde{\omega}_{\rm QM},k_y)=0$ for a given SPI density profile $n_0(x)$: see~\cite{supplemental} for the mathematical details of calculating $D_{\ast}(\tilde{\omega}_{\rm QM},k_y)=0$. In Fig.~\ref{fig:Fig2}(c), we present an example of the complex wavenumber $\tilde{k}_y(\omega) \equiv k_y + i\gamma$ calculated for the real-valued frequencies $\omega > \omega_c$ assuming the SPI in the form of a linearly-varying plasma density ramp extending from $x_{\rm in}=-0$ to $x_{\rm out} = \delta l +0$: $n(x) = n_0(1 - x/\delta l)$ and $\delta l = 0.2 k_p^{-1}$. Note that the resonant point $x_r \equiv x_{\rm UH}$ (where  $x_{\rm UH}/\delta l = \left( \omega_{\rm UH}^2 - \omega^2 \right)/\omega_{p0}^2$) is also real-valued, so the the integration contour $\mathcal{C}$ includes a semi-circle above $x=x_{\rm UH}$. The numerically calculated oscillation (decay) constants $k_y$ ($\gamma$) are plotted in Fig.~\ref{fig:Fig2}(c) as solid-red (dashed-blue) lines, respectively. Note that the decay constant $\gamma$ is normalized to $k_p^2 \delta l$ in Fig.~\ref{fig:Fig2}(c): a result that holds rather accurately for sharp plasma-vacuum interfaces with $\delta l \ll k_p^{-1}$.

\begin{figure}
\centering
\includegraphics[width=1\linewidth]{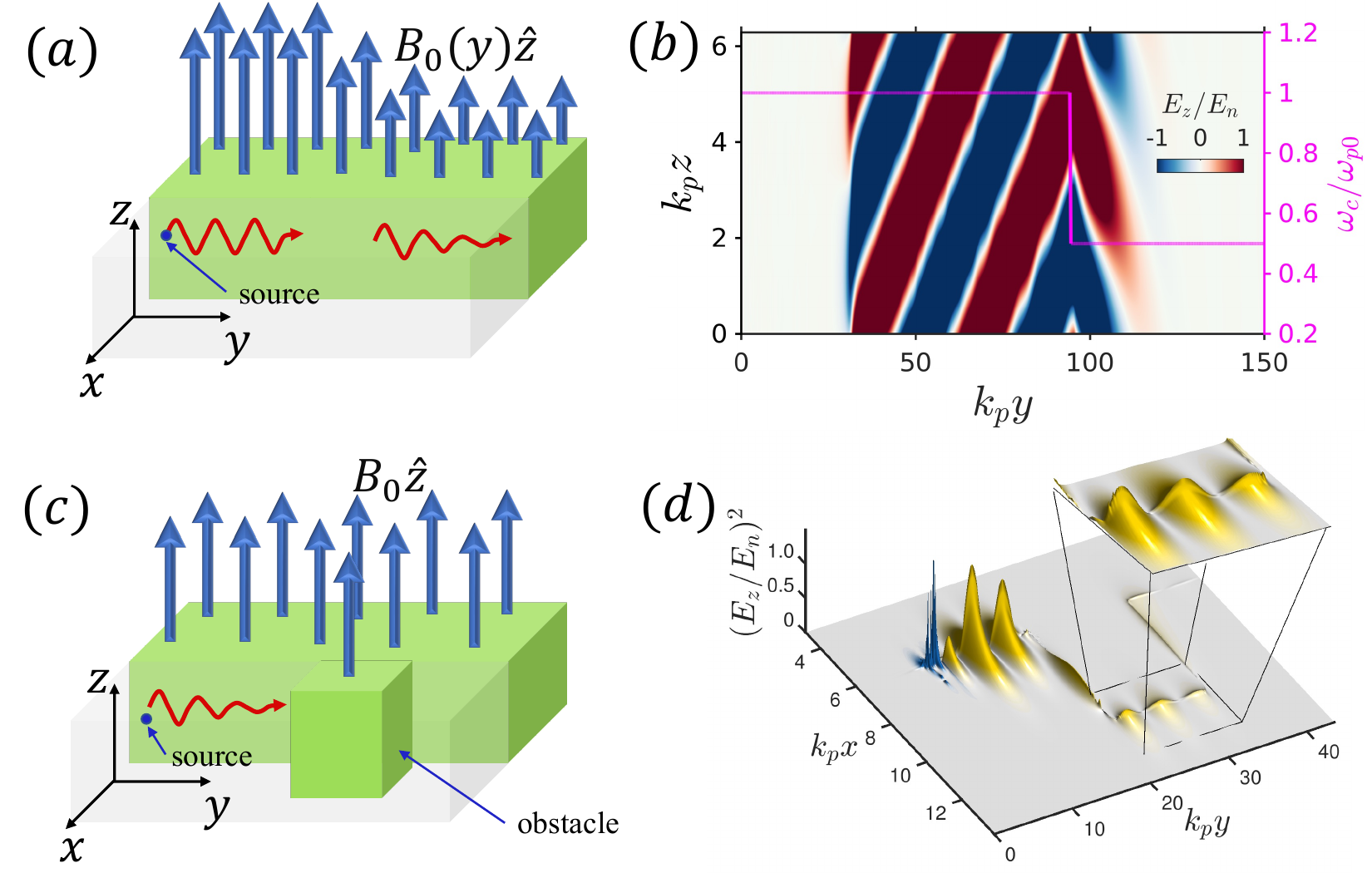}
\caption{\label{fig:Fig3} Reflectionless surface wave propagation in a sheared magnetic field $\vec{B} = B_0 (y) \hat{\rm z}$ (a,b) and around an obstacle (c,d). (b) Refraction of a propagating TSPW ($y < 100k_p^{-1}$) with frequency $\omega_{\rm dr} = 0.67 \omega_{\rm p0}$ into a damped TSPW ($y > 100k_p^{-1}$): $E_z$ at the plasma-vacuum interface (color-coded) and the cyclotron frequency $\omega_c(y)$ (magenta line, scale on the right) linearly sheared on the $L_y^{\rm shear}=2 k_p^{-1}$ scale, where $\omega_c^{\rm (l)} = \omega_{p0}$ and $\omega_c^{\rm (r)} = 0.5 \omega_{p0}$. Rest plasma parameters in (a,b): same as in Fig \ref{fig:Fig1}. (d) A snapshot of time-averaged $E_z^2$ (yellow color) of a damped TSPW with $\omega_{\rm dr} = 0.7 \omega_{\rm p0}$. Plasma parameters in (c,d): $k_{\rm z}/k_{\rm p} = 0.8$, $\omega_{\rm c}/\omega_{\rm p0} = 0.5$, and $k_p \delta l \approx 0.6$. Inset: the TSPW spatially decays inside each of the propagation arms.}
\end{figure}

It has been demonstrated~\cite{Pakniyat_ieee_2020,supplemental} that in the HPI case ($\delta l = 0$) there always exists a real-valued $k_y$ root for the following spectral range: $\omega_{1} < \omega < \omega_{2}$, where $\omega_{2,1} = \left( \sqrt{2 \omega_{p0}^2 + \omega_c^2} \pm \omega_c \right)/2$. For a fixed $k_z/ k_p =1$, the resulting dispersion curve (black solid line) for a HPI-TSPW is plotted in Fig.~\ref{fig:Fig2}(c). In the SPI case as long as $x_{\rm UH}$ is located inside the ramp, the propagation wavenumber $\tilde{k}_y$ must be complex, with its imaginary part $\gamma$ proportional to the fields discontinuity -- just as our plots of $k_y$ and $\gamma$ in Fig.~\ref{fig:Fig2}(c) indeed demonstrate for a specific density ramp with $\delta l = 0.2 k_p^-1$.

Remarkably, while the $k_y(\omega)$ curves for the HPI- and SPI-TSPW overlap for small values of $k_y$, continuum damping removes the unphysical flattening of the HPI-TSPW dispersion curve for large propagation wavenumbers $|k_y| \gg k_p$, thereby removing the divergences of the local density of states and of the thermal electromagnetic energy density~\cite{Archambault_prb_2009,Buddhiraju_nc_2020}.
As a result, the combination of the lossless TSPW ($\omega_{-} < \omega < \omega_{c}$) and the lossy quasi-mode ($\omega_c < \omega < \omega_{+}$) produces a propagation mode spanning the entire topological bandgap bracketed between the propagation bands of the bulk continua and restores BEC principle without introducing temperature \cite{Gangaraj_prl_2020} or small wavelength cutoff\cite{Han_prb_2022}. Importantly, the group velocity $\partial \omega/\partial k_y > 0$ of the mode is positive for all frequencies inside the topological bandgap, thus indicating unidirectional propagation that is expected to be protected against backscattering. The normalized spatial damping rate $\gamma/k_p^2 \delta l$ of the topological quasi-mode is observed to be at its minimum for the long-wavelength modes with positive $k_y < k_p$, and to increase for the slowly-propagating short-wavelength modes.

How ``real" are the continuum-damped quasi-modes, given that they represent only one part of the plasma response, in addition to the continuum of modes from the $\omega_{\rm c} < \omega < \Omega_{\rm UH}$ range~\cite{sedlacek_1971,case_PoF60,briggs_PoF70}? In fact, the {\it long-term} behavior of the fields inside the SPI region is expected to be  non-exponential~\cite{strogatz_prl92,Shvets_pop_1999}. However, their early- and medium-term behavior is accurately described as exponential decay in time or space, depending on the initial/boundary conditions. This is confirmed by our PIC simulations presented in Fig.~\ref{fig:Fig1}(d) that show nearly-exponentially decaying surface waves with the damping constant $\gamma_{\rm PIC} \approx 0.08 k_p$ and the propagation number $k_{y \rm PIC} \approx 0.8 k_p$. Both are in close agreement with the corresponding complex-valued propagation constant of the surface quasi-mode: $\tilde{k}_y = \left( 0.8 - 0.075i\right) k_p$.

Another sign of the quasi-modes physical significance is that PIC simulations indeed confirm their unusual field distribution predicted by the analytic theory. For example, it follows from Eq.(\ref{eq:ExEy}) that the largest field component $E_x \propto 1/\epsilon_t(x)$ is expected to peak and change its sign at the upper-hybrid location $x_{\rm UH}$ inside the linear SPI density ramp ($6 < k_{p} x < 6.2$) as observed in Figs.~\ref{fig:Fig2}(d,e), where it can be seen that $|E_x| \gg |B_z|$ inside the ramp. Moreover, the $B_z(x)$ profile is nearly-discontinuous at $x=x_{\rm UH}$. This is consistent with the jump condition for $\mathbf{\psi}$ that was used to derive the quasi-mode dispersion relation given in  \cite{supplemental}.

Next, we demonstrate the reflectionless conversion of a collisionless TSPW into a weakly-damped quasi-mode. The setup is shown in Fig.~\ref{fig:Fig3}(a), where the boundary between the two (left and right) plasmas with identical density profiles comprises a linear magnetic shear region with the length $L_y = 2k^{-1}_{p}$, and the radiation source on the left side of the shear region launching a TPSW with $\omega_{ \rm dr} = 0.67 \omega_{\rm p0}$. The corresponding cyclotron frequencies on the two sides of the shear region are $\omega_c^{\rm (l)} = \omega_{p0}$ and $\omega_c^{\rm (r)} = 0.5\omega_{p0}$, respectively, and the two bulk plasmas share a common topological bandgap that includes the source frequency: $\omega_{-}^{\rm (l,r)} < \omega_{\rm dr} < \omega_{+}^{\rm (l,r)}$, where $\omega_{-}^{\rm (l)} \approx 0.45$, $\omega_{+}^{\rm (l)} \approx 1 $, $\omega_{-}^{\rm (r)} \approx 0.25$ and $\omega_{+}^{\rm (r)} \approx 1 $ (see Fig.S1 and the details of bulk band structures in~\cite{supplemental}).

Because the source frequency also satisfies the $\omega_c^{\rm (r)}  < \omega_{ \rm dr} < \omega_c^{\rm (l)}$ condition, the launched mode is an undamped (damped) TPSW on the left (right) sides of the magnetic shear region centered at $y=100k_p^{-1}$. The corresponding propagation wavenumbers $k_y$ on the opposite sides of the shear region have opposite signs across the magnetic shear layer: $k_y^{\rm l,r} \approx \mp 0.2 k_p$, respectively. Therefore, while the group velocities $v_y^{(gr)} = \partial \omega/\partial k_y > 0$ of the TSPWs are positive on the two sides of the boundary, their phase velocities $v_y^{(ph)} = \omega/k_y$ change sign. Therefore, the electric field profiles shown in Fig.~\ref{fig:Fig3}(b) reveal reflectionless negative refraction~\cite{pendry_prl00} of the incident undamped TPSW into the damped one. A continuum-damped TSPW can thus be used for depositing energy and momentum into localized regions of the plasma defined by tuning the strength of the external magnetic field, as shown in Figs.~\ref{fig:Fig3}(a,b).

Finally, we demonstrate the robustness of damped SPI-TSPW when encountering a sharp rectangular obstacle (see Fig. \ref{fig:Fig3}(c)) in its path. The dimensions of the obstacle are chosen as $L_x \times L_{\rm y} \times L_{\rm z} = 4.7 k^{-1}_{p} \times 12.6 k^{-1}_{\rm p} \times 7.9 k^{-1}_{\rm p}$. As shown in Fig. \ref{fig:Fig3}(d), the launched TSPW seamlessly propagates around the obstacle without any reflection or radiation into the bulk plasma, thus manifesting topological protection despite their damping~\cite{Davoyan_prl_2013}. The collisionless damping observed before and around the obstacle (see inset of Fig. \ref{fig:Fig3}(d)) is a manifestation of the absorption by localized upper-hybrid resonances.

In summary, we have established the existence of collisionlessly-damped, yet  topologically-protected surface quasi-modes in a magnetized plasma. Using PIC simulations and theoretical calculations, we have identified the damping mechanism through the coupling to a continuum of upper-hybrid resonances localized inside the plasma-vacuum domain wall. While the calculations presented here assume mobile electrons in the background of immobile, future work will address the finite ion mass and the possibility of topological effects in multi-species plasmas.

\begin{acknowledgments}
This work is supported by Air Force of Scientific Research (AFOSR) through Stanford University under MURI Award no. FA9550-21-1-0244. The authors would like to thank Texas Advanced Computing Center (TACC) for providing HPC resources.
\end{acknowledgments}

\bibliography{topological_spw_damping_v3.bib}

\end{document}